\documentclass[12pt,a4paper]{article}

  \usepackage{a4wide}
  \usepackage{latexsym}
  \usepackage{epsf}
  \usepackage{amssymb}
  \usepackage{graphicx}
  \usepackage{amsmath, cite}
  \usepackage{amsmath,amssymb,amsthm}
  \usepackage{verbatim}
  \usepackage{hyperref}

\renewcommand{\d}{\textrm{d}}

\newcommand{\be}{\begin{equation}}
\newcommand{\ee}{\end{equation}}
\newcommand{\beq}{\begin{equation}}
\newcommand{\eeq}{\end{equation}}
\newcommand{\ba}{\begin{eqnarray}}
\newcommand{\ea}{\end{eqnarray}}
\newcommand{\bea}{\begin{eqnarray}}
\newcommand{\eea}{\end{eqnarray}}

\renewcommand{\d}{\textrm{d}}

\begin{document}
\setcounter{page}{0}
\begin{titlepage}
\titlepage

\begin{flushright}
IPhT-t13/064
\end{flushright}
\vspace{0.7cm}

\begin{center}

{\LARGE \bf{The quantization problem in\\\vspace{0.5cm} Scherk--Schwarz compactifications
}}

\vspace{1.1 cm} {\large M.\ Gra\~na,$^a$ R.\ Minasian,$^a$ H.\ Triendl$^a$ and T.\ Van
Riet$^b$}\\

\vspace{0.8 cm}{$^{a }$Institut de Physique Th\'eorique, CEA
Saclay, \\CNRS URA 2306 ,  F-91191 Gif-sur-Yvette, France}\\
\vspace{.1 cm} {$^b$ Instituut voor Theoretische Fysica, K.U. Leuven,\\
Celestijnenlaan 200D B-3001 Leuven, Belgium} \footnote{{\ttfamily{\scriptsize mariana.grana, ruben.minasian,  hagen.triendl @ cea.fr,  thomasvr @ itf.fys.kuleuven.be}}}\\

\vspace{0.8cm}

{\bf Abstract}
\end{center}

\begin{quotation}

We re-examine the quantization of structure constants, or equivalently the choice of  lattice in the so-called flat group reductions, introduced originally by Scherk and Schwarz. Depending on this choice, the vacuum either breaks supersymmetry and lifts certain moduli, or preserves all supercharges and is identical to the one obtained from the torus reduction. Nonetheless the low-energy effective theory proposed originally by Scherk and Schwarz is a gauged supergravity that describes supersymmetry breaking and moduli lifting for all values of the structure constants.  When the vacuum does not break supersymmetry, such a description turns out to be an artifact of the consistent truncation to left-invariant forms as illustrated for the example of ISO(2). We furthermore discuss the construction of flat groups in $d$ dimensions and find that  the Scherk--Schwarz algorithm is exhaustive. A classification of flat groups up to six dimensions and a discussion of all possible lattices is presented.

\end{quotation}
\end{titlepage}

\section{Introduction}

Since the original work by Scherk and Schwarz \cite{Scherk:1978ta, Scherk:1979zr}, truncations onto left-invariant forms on group manifolds (and later cosets \cite{cosetreview1}) have become a standard tool for generating solutions of higher-dimensional gravity, where the geometry is a direct product of a lower-dimensional gravitational background with a manifold that allows for a transitive action by a group $G$.
Restricting to fields that are singlets under the group action (which in our conventions will be taken form the  left) ensures that  the solutions of the lower-dimensional theory solve the higher-dimensional equations of motion. Such reductions are known as consistent truncations.

The groups in question are typically not compact, and it is assumed that they admit a suitable lattice action $\Gamma$ and yield compact manifolds $G/ \Gamma$.\footnote{Note that by abuse of notation, we denote for simplicity the division of the group $G$ by the discrete subgroup $\Gamma$ from the left by $G/ \Gamma$, though notation suggests that the quotient is from the right.} However, consistent truncations to left-invariant forms can also be carried out for non-compact group manifolds, and even for non-compact groups admitting  multiple lattice actions, the truncation is insensitive to the choice of the lattice.\footnote{In \cite{Hull:2005hk} it was conjectured that in string theory different lattices should represent different theories at high energies with the same low-energy dynamics.} Hence there is no reason to believe that in general the consistent truncation captures the full low-energy effective theory. It should rather capture a subset of the light fields. In \cite{Dall'Agata:2005fm} it was pointed out that for certain discrete choices for the structure constants of the group $G$, the Scherk--Schwarz reduction does not represent the full low-energy effective theory of the string compactification.

In the context of supergravity, consistent reductions on group manifolds do not break any supersymmetry of the action. For example, reductions on group manifolds of eleven-dimensional or ten-dimensional type II supergravity generate gauged maximal supergravities in lower dimensions. For phenomenological applications it is desirable to find lower-dimensional vacuum solutions of ten- or eleven-dimensional supergravity that break (part of the) supersymmetry and for which the number of massless fields is as small as possible (ideally none). Such compactifications were found by Scherk and Schwarz in \cite{Scherk:1979zr} using compact group manifolds that admit a flat metric for certain values of the four-dimensional scalar fields in the left-invariant metric.\footnote{We are using the term  Scherk--Schwarz compactifications only for the flat group reductions.} This point in scalar field space then corresponds to a Minkowski vacuum. Away from the vacuum the curvature becomes negative, which implies that those scalars that destroy the flatness property have positive mass. When analyzing the lower-dimensional gauged supergravity, one finds that most scalars in the left-invariant metric are indeed stabilized and that furthermore all supersymmetries have been broken. This presents a clear phenomenological advantage compared to ordinary torus reductions which give rise to supersymmetric Minkowski solutions with the full $\mathcal{N}=8$ multiplet of massless excitations, described by ungauged supergravity.


The aim of this note is to emphasize  the importance of the choice of lattice (or equivalently the quantization of the structure constants).
In particular, for certain choices of the structure constants, the flat group compactifications possess exactly the same supersymmetry and the same massless spectrum as the torus ones. This is due to a simple mathematical fact, proven by Wolf \cite{Wolf}:
\\
\\
{\bf Theorem 1:} \emph{Any Riemannian homogeneous flat space $\mathcal{M}$ is the direct product of the Euclidean plane with the torus: $\mathcal{M} =\mathbb{R}^m \times \mathbb{T}^n$.}\\

This implies in particular that if $\mathcal{M}$ is a compact flat group manifold whose lattice is preserved by the group action, then $\mathcal{M}$ is a torus (as a Riemannian manifold). Therefore at the Scherk--Schwarz vacuum the ten- or eleven-dimensional background is a (flat) torus background. If the gauged supergravity derived from the consistent truncation to left-invariant forms has less massless scalars than the effective action of the torus compactification, this means that some massless scalars of the ten- or eleven-dimensional theory have been truncated away.

At this point one would be tempted to try to argue away the Scherk--Schwarz compactifications as artifacts of truncation.\footnote{Unfortunately, we succumbed to such a temptation in the first version of this paper.} The key notion here is homogeneity. Any manifold of the form $\mathcal{M}= G/ \Gamma$ is homogeneous as a topological manifold, as the action of $G$ from the right is transitive on $\mathcal{M}$. However, the metric on $\mathcal{M}$ is only left-invariant and therefore the action from the right is not isometric.
When the lattice $\Gamma$ does not commute with the generators of the Lie group, the group action from the left on the manifold $\mathcal{M}= G/ \Gamma$ is not well-defined and $\mathcal{M}$ is not homogeneous as a Riemannian manifold (and therefore the theorem does not apply).


In fact, explicit constructions of flat inhomogeneous manifolds exist in the literature, notably in \cite{Porrati, FKPZ}. All flat manifolds have a torus as a covering space, and in some cases a torus acted upon by a torsion-free crystallographic group (rotations and translations) may serve for viable Scherk--Schwarz compactifications. The knowledge about the  crystallographic groups is used for a classification of flat manifolds \cite{Morgan}, which we shall discuss in this work. However, as we will see, many constructions lead to manifolds that are not parallelizable and that are in many cases even non-spin or non-orientable. Some of these arise from higher dimensional algebras quotiented  by a group that has a continuous as well as a discrete piece. The existence of an underlying  $d$-dimensional flat solvable group (with a discrete isotropy) is however the crucial ingredient in order to do a Scherk--Schwarz reduction on a $d$-dimensional manifold, and therefore these cases do not constitute suitable internal manifolds.

Note that though $\mathcal{M}$ might not be homogeneous as a Riemannian manifold, the reduction on left-invariant forms of $G$ can still be well-defined and
independent of the lattice $\Gamma$. To be more precise, it does depend on $\Gamma$ only implicitly -- through the structure constants. As we shall show explicitly in Section \ref{sec:3d},  their quantization rule is determined by the lattice. They, in turn, give rise to the gaugings. In all cases, regardless  of  the internal space $\mathcal{M}$ being  homogeneous or not, the reduction gives rise to a consistent truncation.  Whether this truncation represents the low-energy action is entirely determined by the properties of $\Gamma$.
In the following we will explain this in more detail.

On the group manifold one can build a left-invariant metric, defined as
\be
\d s_{G}^2 = M^{(G)}_{ab} \eta^a  \eta^b\,,
\ee
with $\eta^a$ being the left invariant Maurer--Cartan forms
\be
g^{-1}\d g = \eta^a T_a \, ,
\ee
where $g$ is an arbitrary group element and $T_a$ the set of Lie algebra generators. The bilinear form  $M^{(G)}$ is any symmetric positive definite matrix that does not depend on the coordinates of the manifold. In the context of dimensional reduction, $M^{(G)}$  depends on the external coordinates and contains a set of scalar fields living on the external space. These scalar fields span the coset $GL(n,\mathbb{R})/SO(n)$.  When the physical fluctuations in the vacuum are restricted to these metrics and all the gauge fields are expanded accordingly in the basis of left-invariant forms, the reduced theory is a maximal gauged supergravity. For a subclass of these metrics, the manifold will be flat, but for a generic choice it will be curved. The vacuum and its moduli space are defined by the space of flat metrics.\footnote{The equations of motion require only Ricci-flatness, but Ricci-flat homogeneous parallelizable manifolds are necessarily flat \cite{flat}.}

Whenever the lattice commutes with the group action, the one-forms $\d x^a$ are globally defined and we therefore have the more familiar family of metrics on a torus
\be
\d s_{T}^2 = M^{(T)}_{ab} \d x^a  \d x^b\,,
\ee
where $x^a$ are the usual torus angles. Similarly, $M^{(T)}_{ab}$ describes a $GL(n,\mathbb{R})/SO(n)$ coset of lower-dimensional scalar fields.
The two sets of metrics coincide in those metrics that are simultaneously flat and left-invariant.

When  the rest of the fields of the higher-dimensional supergravity is expanded in terms of  $\eta^a$ resp.~$\d x^a$, the lower-dimensional action is gauged resp.\ ungauged maximal supergravity. The reason the vacuum of the lower-dimensional gauged supergravity seems to break supersymmetry is simply because the Killing spinors of the higher-dimensional vacuum (e.g.\ Mink$_4 \times \mathbb{T}^7$) are not left-invariant under $G$.
So which reduction should be chosen? The answer depends on the purpose of the dimensional reduction. When used as a solution-generating technique, the left-invariant truncation and its associated gauged supergravity will allow one to find non-trivial solutions (with non-constant scalars) beyond the usual flat torus solutions. However when  interested in the low-energy effective theory, one needs to describe the lightest excitations. These are obviously given by the second set of metrics, which break the left-invariant symmetry.
A similar problem arises  when counting  the amount of supersymmetry in the ten-dimensional vacuum \cite{Grana:2006kf}.  Clearly,  the gauged supergravity analysis can only capture left-invariant supercharges and so generically misses the majority of conserved supersymmetries.

This dichotomy could happily be avoided for the groups for which the corresponding Lie algebra cohomology is isomorphic to the de Rham cohomology on the compact manifold $G/\Gamma$ obtained by the lattice action on the group. The isomorphism holds for the so-called completely solvable groups, a class that includes all nilpotent groups. In general however there exists only an inclusion map between the respective cohomologies, and the Betti numbers for the Lie algebra cohomology give only a lower bound for the corresponding numbers for de Rham
cohomology (see for instance \cite{Bock} for more details). In particular, flat groups are never completely solvable.

In the following section we give an explicit example that illustrates these facts, based on the unique three-dimensional flat group manifold ISO(2).  A higher-dimensional version of this example can be found in \cite{Dall'Agata:2005fm}, where it was already noted that for certain discrete parameters the theory is indeed represented by the flat torus reduction. Here we show all the discrete parameters that allow a lattice action and therefore a compactification, and distinguish between those giving rise to homogeneous manifolds (and therefore tori), and those that do not. The Scherk--Schwarz reductions (for flat groups) only represent the low-energy effective action in the latter case.

In Section \ref{sec:higherd} we present all the higher dimensional flat solvable algebras. We point out that the construction proposed by
Scherk and Schwarz is exhaustive due to a theorem by Milnor \cite{Milnor}. The corresponding lattices were also classified algorithmically \cite{Morgan} and constructed explicitly up to dimension four and partially in dimension five. We present a very simple description of these lattices for all dimensions, and show explicitly a few examples in four dimensions. We furthermore complete the classification in five dimensions, and discuss quantization conditions in dimension six.

\section{Three-dimensional solvmanifolds}
\label{sec:3d}
\subsection{The solvable group ISO(2) and maximal gauged supergravity}
For any group manifold the left-invariant Maurer-Cartan forms satisfy
\begin{equation}
\d\eta^a=-\frac{1}{2} f^a_{bc} \, \eta^b \wedge \eta^c \ ,
\end{equation}
where $f^a_{bc}$ are the structure constants of the associated Lie algebra.
Any traceless (unimodular) Lie algebra in three dimensions can be written as
\begin{equation}
f_{bc}^{a} = \varepsilon_{bcd}Q^{ad}
\end{equation}
with $Q$ diagonal. If we take $Q$ to be rank two and with two positive eigenvalues we get ISO(2):
\begin{equation}
Q = \begin{pmatrix} 0 & 0 & 0 \\
0 & q_1 & 0\\
0 & 0 & q_2
\end{pmatrix}\,,\qquad q_1, q_2>0\,.
\end{equation}
The left invariant Maurer--Cartan forms, $\eta^a$ therefore  obey
\begin{equation}\label{algebra}
\d\eta^1 = 0\,,\qquad \d\eta^2 =   q_1\, \eta^1\wedge\eta^3\,,\qquad \d\eta^3 = -q_2 \, \eta^1\wedge\eta^2\,.
\end{equation}
A (local) coordinate representation can be found as follows ($(x_1, x_2, x_3)$ are real coordinates on ISO(2))
\begin{align}
& \eta_1 = \d x^1 \,, \\
& \eta_2 = \cos(\sqrt{q_1q_2}x^1) \d x^2 + \sqrt{\frac{q_1}{q_2}}\sin(\sqrt{q_1q_2} x^1)\d x^3\,, \label{eta2}\\
& \eta_3 = -\sqrt{\frac{q_2}{q_1}}\sin(\sqrt{q_1q_2}x^1) \d x^2 +  \cos(\sqrt{q_1q_2} x^1)\d x^3\,. \label{eta3}
\end{align}
For notational simplicity the, we shall take
\be \label{same_q}
q_1=q_2=q\, .
\ee
This can always be realized by rescaling $\eta^3$ with respect to $\eta^2$ (i.e.\ rescaling $x^3$ with respect to $x^2$). The Lie algebra is solvable of degree one. The associated group manifold is therefore called a solvable manifold.

The cohomology of the left-invariant $p$-forms, denoted with Betti numbers $b_p^L$ is
\begin{equation}
b_0^L = 1 \,,\quad b_1^L = 1 \,,\quad b_2^L=1 \,, \quad b_3^L =1 \,,
\end{equation}
where the generator of the first cohomology is $\eta_1$.
This can differ from the dimensions of the de Rham cohomology groups, as we will see, but it always gives a lower bound on the de Rham Betti numbers. Before we further discuss this and the related issue of compactness, let us describe the reduction to maximal gauged supergravity.

The left-invariant metric on this manifold is given by
\begin{equation}\label{metric_leftinv}
\d s_{G}^2 = M^{(G)}_{ab} \eta^a  \eta^b\,,
\end{equation}
where $M^{(G)}$ is any symmetric positive definite matrix and hence inside $GL(3, R)/SO(3)$. The curvature vanishes for the following four-dimensional family of metrics\footnote{Note that the parameters $c$ and $d$ can be absorbed into the definition of $\eta^2$ and $\eta^3$, by defining the equivalent left-invariant one-forms $\tilde \eta^2 = \eta^2 +\tfrac{c}{b} \eta^1$ and $\tilde \eta^3 = \eta^3 +\tfrac{d}{b} \eta^1$, which also fulfill \eqref{eta2} and \eqref{eta3}. Therefore, the parameters $c$ and $d$ correspond to a different choice of left-invariant one-forms.}
\begin{equation} \label{ab}
M^{(G)} = \begin{pmatrix} a+\tfrac{c^2+d^2}{b}  & c & d \\
c & b & 0 \\
d & 0 & b
\end{pmatrix}\,,\qquad a,b>0\,.
 \end{equation}

Finally we briefly review the rough structure of the compactified theory
in either case, i.e.\ when the fields are expanded in the left-invariant basis, and when the fields are expanded in the one-forms $\d x^a$.
The latter is only possible when the one-forms are globally defined, which is the case only for certain quantization conditions, as we will see in detail later.  We will focus on the reduction of the metric sector, the dilaton and form fields can be worked out accordingly. When the fields are expanded in the left-invariant basis, the scalar potential and its supersymmetric solutions have been worked out in \cite{AlonsoAlberca:2003jq}. If we start in ten dimensions then the metric ansatz is
\begin{equation}
\d s^2_{10} = {\rm e}^{2\alpha v}\d s_{7}^2 + {\rm e}^{2\beta
v}M_{ab} (\theta^a +A^a) (\theta^b+A^b) \,.
\end{equation}
with $\theta^a$ either being $\eta^a$ or $\d x^a$, $A^a$ are Kaluza-Klein vectors and
the numbers $\alpha, \beta$ chosen such that we
end up in lower-dimensional Einstein frame with canonically
normalized fields: $\beta =-5\alpha/3$ and $\alpha^2 =3/80$.
The reduced action is
\begin{equation}
S = \int \sqrt{-g}\Bigl( R - \tfrac{1}{2}(\partial v)^2+ \tfrac{1}{4} \text{Tr}(M^{-1} (D M) M^{-1} (D M) ) -\tfrac14 {\rm e}^{-2(\alpha-\beta)v} M_{ab} F^a  \cdot F^b - V\Bigr)\, ,
\end{equation}
where we have for the truncation to left-invariant modes $M= M^{(G)}$ and
\begin{equation}
D M^{(G)}_{ab} = \partial M^{(G)}_{ab} + 2 f^d_{c(a} M^{(G)}_{b)d}  A_{(G)}^c  \, , \quad F^a_{(G)}=d A^a_{(G)} + 2 f^a_{bc} A^b_{(G)} \wedge A^c_{(G)} \, .
\end{equation}
The scalar potential in seven dimensions can be written as
\begin{equation}
V =\tfrac{1}{2} {\rm e}^{2(\alpha-\beta)v} [2 \text{Tr}(QM^{(G)}QM^{(G)}) - \text{Tr}(QM^{(G)})^2] \,.
\end{equation}
In the example of ISO(2), Eq.\ \eqref{algebra} gives $Q={\rm diag}(0,q,q)$ so that $V$ gives a mass to $M^{(G)}_{23}$ and $M^{(G)}_{22} - M^{(G)}_{33}$.
Furthermore, the scalars $M^{(G)}_{12}$ and $M^{(G)}_{13}$ are eaten by the Kaluza-Klein vectors $A_{(G)}^2$ and $A_{(G)}^3$, which become in turn massive. Thus the Minkowski vacuum is indeed parameterized by the family \eqref{ab}, where $M^{(G)}_{12}$ and $M^{(G)}_{13}$ just denote physically equivalent choices of left-invariant one-forms and the massless fields $M^{(G)}_{11}$, $M^{(G)}_{22}+M^{(G)}_{33}$
and $A_{(G)}^1$ would coincide with their counterparts on the torus, if the metric $M^{(T)}$ can be defined. The massive modes $M^{(G)}_{23}$ and $M^{(G)}_{22} - M^{(G)}_{33}$ as well as the massive vectors $A_{(G)}^2$ and $A_{(G)}^3$ however would correspond to the first massive states in the Kaluza-Klein tower of the torus. Whether the massless states of the torus are there or not determines if the truncation to left-invariant modes represents the low energy limit of the compactification. As we shall discuss now, the answer to this question  depends on the choice of lattice.

\subsection{Lattices for ISO(2)}

The group ISO(2) admits several lattices, whose action yields compact manifolds \cite{Grana:2006kf,Bock}. Any such resulting manifold  $G/\Gamma$ is called a solvmanifold.

 For the simplest class of equivalent lattices the coordinates $x^2$ and $x^3$ can, without loss of generality, be chosen to have unit length, i.e.\ we quotient by the usual torus lattice so that $x^2 \sim x^2 +m^2$ and $x^3 \sim x^3  + m^3$ for integers $m^2$ and $m^3$. The identification involving the coordinate $x^1$ is more subtle. In general one can take
\begin{equation}\label{identISO2} \begin{aligned}
 x^1 \to & \,\, x^1 + m^1\, , \\
 x^2 \to & \,\, \cos(qm^1) x^2  - \sin(qm^1) x^3 \, , \\
 x^3 \to & \,\,  \sin(qm^1) x^2 + \cos(qm^1) x^3 \, ,
\end{aligned}\end{equation}
with integer $m^1$. The identification \eqref{identISO2} leaves the one-forms $\eta^i$, $i=1,2,3$, invariant, but it cannot be defined for arbitrary $q$. In other words, $q$ obeys a quantization rule, which for the simplest class of lattices reads
\begin{equation}\label{quant2}
q= 2 \pi k  \, ,
\end{equation}
for some integer $k$. Such a lattice does not break the group ISO(2), i.e.\ $\mathcal{M}=G/\Gamma$ is homogeneous as a Riemannian manifold. One the other hand, the identification \eqref{identISO2} just gives the torus lattice, in agreement with Wolf's theorem. This can be seen by the fact that $x^2$ and $x^3$ also define globally defined one-forms $\d x^2$ and $\d x^3$. Therefore, the de Rham cohomology groups are the ones of the three-torus ($b_i=(1,3,3,1)$) and larger than the Lie cohomology of ISO(2) such that the consistent truncation to left-invariant forms misses some massless fields in this case.
In other words we could now study the family of globally-defined flat torus metrics
\begin{equation}
\d s_{T}^2 = M^{(T)}_{ab} \d x^a  \d x^b\,,
\end{equation}
with $M^{(T)}$ being any symmetric positive definite matrix (living in $GL(3, R)/SO(3)$). A generic metric in this family does not possess the original $ISO(2)$ symmetry. If one reduces now on the one-forms $\d x^a$, we find that there are no gaugings in four dimensions, i.e.\ $D M^{(T)} = \partial M^{(T)}$, $F^a_{(T)}=dA^a_{(T)}$, and the consistent truncation leads to maximal ungauged supergravity with a maximally supersymmetric vacuum.

For further clarification let us now discuss supersymmetry for the ISO(2) truncation. The left-invariant spinors on the internal space are
\begin{equation} \label{spinors_ISO2}
 \epsilon_{(G)}^1 =\begin{pmatrix} 1 \\ 0 \end{pmatrix} \ , \qquad  \epsilon_{(G)}^2 =\begin{pmatrix} 0 \\ 1  \end{pmatrix} \, ,
\end{equation}
for the vielbein $\eta^i$. These spinors are not covariantly constant, since the connection has a non-trivial component $\omega^2{}_3 = - q \eta^1$.
Note that these spinors are related to the left-invariant one-forms by
\begin{equation}
 \eta^a = \sigma^a_{ij} \bar \epsilon_{(G)}^j \sigma_b \epsilon_{(G)}^i \d x^b \, .
\end{equation}
On the other hand, for the same vielbein the spinors
\begin{equation} \label{spinors_torus}
 \epsilon_{(G)}^1 =\begin{pmatrix} \cos(\tfrac12 q x^1) \\ i \sin(\tfrac12 q x^1) \end{pmatrix} \ , \qquad  \epsilon_{(G)}^2 =\begin{pmatrix} i \sin( \tfrac12 qx^1) \\ \cos( \tfrac12 q x^1)  \end{pmatrix} \, ,
\end{equation}
are covariantly constant. In the Scherk--Schwarz reduction to left invariant modes they are truncated away. However, for the lattice defined by \eqref{identISO2},\eqref{quant2}, these spinors are well-defined. This shows that the Minkowski vacuum preserves 32 supercharges in the ten-dimensional theory. From the torus point of view, the spinors \eqref{spinors_torus} are the massless modes, while the spinors \eqref{spinors_ISO2} are (massive) Kaluza-Klein spinor modes.

The situation is different for lattices $\Gamma$ that lead to inhomogeneous solvmanifolds $\mathcal{M}=G/\Gamma$. There are four topologically distinct cases \cite{Morgan}. The first two classes of lattices still have an identification of the form \eqref{identISO2} for $x^1$, but now the quantization condition \eqref{quant2} is changed to
\begin{equation}\label{quant_inhom}
q= 2 \pi k + 2 \pi/n \, ,
\end{equation}
where $n=2,4$ labels the two topologically distinct cases. Under the identification \eqref{identISO2} the generators of ISO(2) that shift $x^2$ or $x^3$ are rotated into each other for $n=4$ (and each inverted for $n=2$), so that they are not well-defined any more on the solvmanifold.\footnote{In other words the lattice does not commute with the generators of the Lie group. Only if the lattice is in the center of the group, the manifold can be homogeneous (as a Riemannian manifold). This is only true for the torus, as stated by Theorem~1.} Therefore the manifold does not allow for any transitive group action from the left and thus is not homogeneous (as a Riemannian manifold) any more. In this way it evades Theorem~1.

The other two topologically distinct solvmanifolds have the identifications
\begin{equation}\label{identZ36} \begin{aligned}
 x^1 \to & \,\, x^1 + m^1\, , \\
 x^2 \to & \,\, \cos(qm^1) x^2  - \sin(qm^1) x^3 + m^2 - \tfrac{1}{2} m^3\, , \\
 x^3 \to &\,\,  \sin(qm^1) x^2 + \cos(qm^1) x^3 + \tfrac{\sqrt{3}}{2}  m^3\, ,
\end{aligned}\end{equation}
with $m^i$ being integers and the quantization condition in \eqref{quant_inhom} now being $n=3$ and $n=6$ for the two topologically distinct manifolds. Once more, \eqref{identZ36} does not leave the generators of ISO(2) invariant, therefore yielding an inhomogeneous manifold.
In \eqref{identZ36} the lattice spanned by $m^2$ and $m^3$ is hexagonal, to reflect the invariance under $\mathbb{Z}_6$. Alternatively, one can take linear combinations of $x^u$, $u=2,3$, to bring \eqref{identZ36} into the integral form
\begin{equation}\label{identZ36INT} \begin{aligned}
 x^1 \to & \,\,  x^1 + m^1\, , \\
 x^u \to & \,\, (B_n^{m^1})^u{}_v x^v   + m^u  \, ,
\end{aligned}\end{equation}
where we defined
\begin{equation}\label{Bn36}
 B_n = (-1)^n \left( \begin{aligned} 0 && -1 \\ 1 && 1 \end{aligned} \right) \, .
\end{equation}

Note that each of these four solvmanifolds can be obtained from the torus by taking a freely acting $\mathbb{Z}_n$ quotient. This $\mathbb{Z}_n$ preserves the left-invariant one-forms and therefore, one can still use the ISO(2) reduction of Scherk and Schwarz on the group to obtain a maximal gauged supergravity, which is then preserved by the $\mathbb{Z}_n$ quotient (the $\mathbb{Z}_4$ case is discussed in detail in \cite{Porrati, FKPZ}). However, here the one-forms $\d x^2$, $\d x^3$ are not globally defined, or in other words they are not preserved by the quotient and are therefore projected out from the spectrum. Moreover, the covariantly constant spinors \eqref{spinors_torus} are not well-defined either on the inhomogeneous solvmanifold (i.e.\ for \eqref{quant_inhom}). Therefore, the ten-dimensional Minkowski vacuum is non-supersymmetric and the Betti numbers of the solvmanifold coincide with the Betti numbers of group cohomology (and are one). From the torus perspective, the $\mathbb{Z}_4$ action requires that the two-torus in the $x^2$-$x^3$ plane to be rectangular, while the $\mathbb{Z}_3$ and $\mathbb{Z}_6$ action require the torus to span an angle that is a multiple of $\pi/3$, and therefore the modulus corresponding to its complex structure is truly fixed. An exception is the $\mathbb{Z}_2$ action, which still projects out the Killing spinors so that the vacuum is non-supersymmetric, but does not fix the complex structure of the fibre torus.

Apart from the cases discussed so far there are two other three-dimensional flat solvmanifolds, which have first Betti number equal to two \cite{Morgan}. Both admit a one-dimensional circle fiber which is inverted when going around a one-cycle in the torus base. One of them is just the Klein bottle times a circle, while for the other example the Klein bottle is fibred over the circle \cite{Auslander2}. Therefore both examples are non-orientable, thus not parallelizable, and no Scherk--Schwarz reduction is possible.

The lesson we learn for the torus quantization \eqref{quant2} has also some consequences for the inhomogeneous solvmanifolds. In particular, if we take in \eqref{quant_inhom} $k$ to be non-zero, the $N=8$ gauged supergravity will not include the lightest massive modes of the background, but only the higher Kaluza-Klein states. Therefore, the physically reasonable quantization charges for the four non-trivial solvmanifolds are
\begin{equation}\label{quantphys}
 q = 2 \pi/n \, ,
\end{equation}
for $n=2,3,4,6$. As discussed above, also the $n=2$ case misses some massless fields in the Scherk--Schwarz reduction.

Interestingly, the above quantisation condition coincides with the quantisation conditions for S-duality transformations in IIB supergravity \cite{Bergshoeff:2002mb}. This comes about as follows. Maximal supergravity in nine dimensions has an $SL(2,\mathbb{R})$ symmetry. If we reduce this theory over a circle and at the same time perform a Scherk-Schwarz  $SL(2,\mathbb{R})$-twist, we obtain 8-dimensional gauged SUGRA. If the twist corresponds to $SO(2) \in  SL(2,\mathbb{R})$, then the gauged supergravity can be seen as a obtained from an $ISO(2)$ compactification of 11-dimensional supergravity. Alternatively, one can regard the  $SL(2,\mathbb{R})$ in nine dimensions as inherited from IIB S-duality. The quantisation of the S-duality symmetry then coincides with the geometric quantisation coming from the $ISO(2)$ lattice described above.

\section{Higher-dimensional solvmanifolds} \label{sec:higherd}

It is natural to ask about the higher-dimensional generalizations.
It turns out that all flat groups are solvable and the associated manifolds are \emph{flat solvmanifolds}. A classification technique for their possible lattices has been found in \cite{Morgan}. As for the tree-dimensional case we shall first discuss the flat groups and then the possible lattice actions.

\subsection{The classification of flat groups} \label{sec:classflat}

A classification of flat groups exists; for it one can consult for instance \cite{2010arXiv1009.3283J}.  As it turns out, this classification is based  on a technique, identical to the prescription  given by Scherk and Schwarz in \cite{Scherk:1978ta,Scherk:1979zr}. As a consequence, one may prove that  the method suggested by Scherk and Schwarz is in fact exhaustive.  We shall now review some relevant facts. The classification of flat groups relies on a theorem by Milnor \cite{Milnor}.
\\ \\
{\bf Theorem 2:} \emph{A Riemannian Lie Group G is flat if and only if its Lie algebra $g$ (endowed with an inner product) splits as an orthogonal direct sum $g = a\oplus n$, where $n$ is an Abelian ideal (the nilradical) and $a$ is an Abelian sub algebra such that $ad X$ is anti-symmetric for $X \in a$ }.\\

In other words, the Lie group consists of translations generated by $X_{a} \in n$ which are rotated into each other by transformations generated by elements $X_u \in a$.
From Theorem 2 we see that the Lie algebra must be of the form
\begin{align}
[X_a, X_b] =0 = [X_u, X_v]\,,\\
[X_a, X_u] = [T_a]_u{}^v X_v\,,
\end{align}
where $T_a^T = -T_a$ for all $a$ and they are sometimes known as the `twist matrices'. This algebra automatically satisfies the Jacobi identities and is solvable.
The associated group manifold is flat when endowed with the unit metric if expressed in terms of the associated Maurer--Cartan forms.
The $X_a$ and $X_u$ form respectively the base and fiber of the Scherk--Schwarz construction.  .

In what follows we put this construction in a practical context and classify the algebras up to dimension 6. For that purpose it is useful to recall the  normal form of an anti-symmetric matrix $T$. The rank $r$ is necessarily even-dimensional and the normal form is block diagonal with $r/2$  blocks of the kind
\be
\begin{pmatrix} 0 & -1 \\ 1 & 0 \end{pmatrix}
 \ee
and the rest is zero.

In three dimensions the algebra of ISO(2) is obviously the unique (non-Abelian) Lie algebra of a flat group since the nilradical must be two-dimensional and the single generator in the complement $a$ acts as a rotation on $n$.
This also exhausts all possibilities in four dimensions. First consider the case the nilradical is three-dimensional. Due to the normal form of a $3 \times 3$ anti-symmetric matrix this algebra must be $ISO(2)\times U(1)$. Then consider the case the nil-radical is two-dimensional. We find only the same case again since there is only (up to rescalings) a single  anti-symmetric $2\times 2$ matrix.

In five dimensions we find more possibilities. First consider the case of the  four-dimensional nilradical. There are two algebras depending on the rank of the single anti-symmetric matrix. If its rank is two, the algebra is $ISO(2) \times U(1)^2$. If its rank is four, this defines a new algebra
\be \label{5dgroup}
[X_1, X_u] = T_u{}^v X_v \,,\qquad [X_u, X_v]=0\,,
\ee
where we have
\be
 T=
\begin{pmatrix}
0 & -1 & 0 & 0 \\
1 & 0 & 0 & 0 \\
0 & 0 & 0 & -1 \\
0 & 0 & 1 & 0 \\
\end{pmatrix}\,.
\ee
For any nilradical of smaller dimension we again find only the algebra $ISO(2) \times U(1)^2$.

In six dimensions, we again find a new algebra. First assume the nilradical to be five-dimensional. If $T$ has rank four or two we have the above two algebras plus an Abelian direction. If the nilradical is however four-dimensional there is a new example
\be \begin{aligned}
& [X_1, X_u] = [T_1]_u{}^v X_v \,,\qquad [X_u, X_v]=0 \, ,\\
& [X_2, X_u] = [T_2]_u{}^v X_v \,,\qquad [X_1, X_2]=0 \, ,
\end{aligned}
\ee
where we defined
\be
 T_1=
\begin{pmatrix}
0 & -1 & 0 & 0 \\
1 & 0 & 0 & 0 \\
0 & 0 & 0 & 0 \\
0 & 0 & 0 & 0 \\
\end{pmatrix}\,,\qquad T_2 = \begin{pmatrix}
0 & 0 & 0 & 0 \\
0 & 0 & 0 & 0 \\
0 & 0 & 0 & -1 \\
0 & 0 & 1 & 0 \\
\end{pmatrix}\,.
\ee
This is nothing but $ISO(2) \times ISO(2)$. Again for smaller nilradicals we do not generate new examples.

With Milnor's Theorem this classification is easily extended to higher dimensions. In the following we will however focus onto the more difficult question, namely which lattices and therefore charge quantizations exist for these groups.

\subsection{The lattices of flat groups}
After discussing the solvable groups of dimension six or less, let us now discuss their possible lattices. A classification of lattices in dimensions three and four as well as a partial classification in dimension five has been performed in \cite{Morgan}. In the following we present the lattices and the corresponding solvmanifolds.

A general lattice identifies the group coordinates of a flat solvable manifold of dimension $n_b+n_f$ as~\cite{Morgan}
\begin{equation} \label{identMorgan} \begin{aligned}
x^a \to & x^a + m^a \, ,\\
x^u \to & (\prod_a (B_a)^{m^a})^u{}_v x^v + m^a M^u_a + s m^u \, ,
\end{aligned}\end{equation}
where $x^a$, $a=1,\dots, n_b$, are identified as the coordinates on the base torus and $x^u$, $u=1,\dots, n_f$, as those on the ``fibre torus'' and $m^a$ and $m^u$ are integers that parameterize the lattice.\footnote{All compact solvmanifolds are nilmanifolds fibered over tori.} Moreover, the lattice is specified by $(B_a, M^u_a, s)$ where the $B_a$ span a finite Abelian subgroup of $Gl(n_f,\mathbb{Z})$.

The form \eqref{identMorgan} is different from the one used for the classification of solvable groups. There the rotations of the fibre preserved lengths and therefore are elements in SO($n_f$), while for the lattices it is rather convenient to assume a rectangular lattice where the linear transformations are not pure rotations, but elements in $Gl(n_f,\mathbb{Z})$. For instance \eqref{identZ36} represents  the frame where the fibre is only twisted by rotations, while \eqref{identZ36INT} corresponds to the frame of \eqref{identMorgan}.

Note that the $B_a$ have determinant one or minus one. To guarantee orientability, the $B_a$ should be within $Sl(n_f,\mathbb{Z})$. $M_a$ are vectors with integer entries and $s$ is a (non-vanishing) integer.
The simplest example of a non-orientable solvmanifold is the Klein bottle in two dimensions, where $B_1=-1$ inverts the fibre. We saw that in three dimensions all non-orientable solvmanifolds are straight-forward generalizations of the Klein bottle. In four dimensions and higher, when the fibre is at least three-dimensional, the non-orientable solvmanifolds become richer, as the $B_a$ combine an involution in some direction and some $\mathbb{Z}_n$ action in the others. An example of a non-orientable solvmanifold in four dimensions has
\begin{equation}
 B_1 =  \left( \begin{aligned} - 1 && 0 && 0\\ 0 && 0 && 1\\ 0 && -1 && 0\end{aligned} \right)\,  \, ,
\end{equation}
which generates a $\mathbb{Z}_4$ subgroup of $Gl(3,\mathbb{Z})$. In general, the $B_a \in Gl(n_f,\mathbb{Z})$ generating non-orientable groups can be understood as composed of an involutions in one direction and generators of an Abelian subgroup in $Gl(n_f-1,\mathbb{Z})$.

Non-orientable solvmanifolds are a subclass of the solvmanifolds that are not parallelizable. If a solvmanifold is not parallelizable, it cannot be of the form $G/ \Gamma$ for a lattice $\Gamma$. Such solvmanifolds are constructed from higher-dimensional groups (see for instance \cite{Auslander2}) and do not allow for a standard Scherk--Schwarz reduction to gauged $N=8$ supergravity.
An example of a five-dimensional orientable but non-parallelizable manifold has been given in \cite{Auslander} and is of the form \eqref{identMorgan} with a two-dimensional base, where $M_a=0$, $s=1$ and
\begin{equation}
 B_1 =  \left( \begin{aligned} 1 && 0 && 0\\ 0 && -1 && 0\\ 0 && 0 && -1\end{aligned} \right)\,, \qquad  B_2 =  \left( \begin{aligned} -1 && 0 && 0\\ 0 && -1 && 0\\ 0 && 0 && 1\end{aligned} \right) \, .
\end{equation}
This manifold is not parallelizable and is not spin, as discussed for example in \cite{Auslander}. In the following we will restrict ourselves to the parallelizable cases in \cite{Morgan} that allow for a Scherk--Schwarz reduction.

In four dimensions, all non-trivial parallelizable examples have a two-dimensional fibre and are therefore very similar to the three-dimensional solvmanifolds we discussed in Section \ref{sec:3d}. If $M_a=0$ (and $s=1$), the manifold is indeed just a three-dimensional solvmanifold times a circle. The cases with non-zero $M_a$ (and $s \neq 1$) instead give fibrations of three-dimensional solvmanifolds over a circle. They have the same Betti numbers and Scherk--Schwarz reductions (and the same charge quantizations) as the cases with $M_a=0$ and therefore give the same maximal gauged supergravity. However, the string spectrum on these spaces can differ.

In five dimensions, we again find the examples that lift from three dimensions and might involve some additional fibrations with $M_a \neq 0$ (and $s \neq 1$)  which have first Betti number equal to three. However, there are also new examples with first Betti number equal to one, which have not been classified in \cite{Morgan}.
For these cases, the base is one-dimensional and $M_1=0$. Therefore, these solvmanifolds are completely defined by the matrix $B_1$ that generates a finite subgroup of $Sl(4,\mathbb{Z})$.
The finite subgroups of $Gl(4)$ have been classified in \cite{crystallography}.
For the discussion of five-dimensional solvmanifolds of first Betti number $b_1=1$ it suffices to consider only cyclic groups, as there is only one rotation $B_1$ in the four-dimensional fibre. Furthermore, if $B$ has an eigenvector with eigenvalue one, this solvmanifold has Betti number larger than one and therefore has already been discussed in \cite{Morgan}. In Table~\ref{ta:cyclic} we give the relevant cyclic subgroups of $Sl(4)$ and possible embeddings of their generator into $Sl(4)$.
\begin{table}
\begin{center}
\begin{tabular}{|c|l||c|l|}\hline
group & \hspace{7pt}  generator & group & \hspace{7pt}  generator \\
\hline &&& \\
$\mathbb{Z}_2$ &  $I' =  \left( \begin{aligned} -1 && 0 && 0&& 0 \\ 0 && -1 && 0&& 0\\ 0 && 0 && -1&& 0\\ 0 && 0&& 0 && -1\end{aligned} \right)$ &
$\mathbb{Z}_8$ & $A_1 = \left( \begin{aligned} 0 && 0 && -1&& 0\\ 0 && 0&& 0 && -1 \\ 0 && 1 && 0&& 0\\ -1 && 0 && 0&& 0\end{aligned} \right)$\\&&&\\
 \hline &&&\\
$\mathbb{Z}_{12}$ & $C_1 = \left( \begin{aligned} 0 && 0 && 1&& 1\\ 0 && 0&& 0 && -1 \\ 0 && 1 && 0&& 0\\ 1 && 0 && 0&& 0\end{aligned} \right)$ &
$\mathbb{Z}_6$ & $K_{11'} = \left( \begin{aligned} 1 && 0 && 0&& -1\\ 1 && -1&& 0 && -1 \\ 0 && 0 && -1&& -1\\ 1 && 0 && 0&& 0\end{aligned} \right)$\\&&&\\
  \hline &&&\\
$\mathbb{Z}_{10}$ & $L_{1} = \left( \begin{aligned} 1 && 1 && 1&& 1\\ -1 && 0&& 0 && 0 \\ 0 && -1 && 0&& 0\\ 0 && 0 && -1&& 0\end{aligned} \right)$&
$\mathbb{Z}_{6}$ & $S_1 = \left( \begin{aligned} 1 && 1 && 0&& 0\\ -1 && 0&& 0 && 0 \\ 0 && 0 && 0&& -1\\ 0 && 0 && 1&& 1\end{aligned} \right)$
\\&&&\\
$\mathbb{Z}_{4}$ & $D_1 = \left( \begin{aligned} 0 && 1 && 0&& 0\\ -1 && 0&& 0 && 0 \\ 0 && 0 && 0&& 1\\ 0 && 0 && -1&& 0\end{aligned} \right)$ &&
$S_2 = \left( \begin{aligned} 1 && 0 && 0&& 1\\ 0 && 0&& -1 && 0 \\ 0 && 1 && 1 && 0\\-1 && 0 && 0&& 0 \end{aligned} \right)$
\\&&&\\
& $D_2 = \left( \begin{aligned} 0 && 0 && 1&& 0\\ 0 && 0&& 0 && -1 \\ -1 && 0 && 0 && 0\\0 && 1 && 0&& 0 \end{aligned} \right)$ &&
$S_3 = \left( \begin{aligned} 1 && 0 && 1&& 0\\ 0 && 0&& 0 && -1 \\ -1 && 0 &&0 && 0\\0 && 1 && 0&& 1 \end{aligned} \right)$
\\&&&\\
 \hline
\end{tabular}
\caption{  \label{ta:cyclic}
Generators of cyclic subgroups of $SL(4)$ that have no eigenvector with eigenvalue one; they are labeled following the nomenclature of \cite{crystallography}. In some cases there are multiple, equivalent generators.}
\end{center}
\end{table}
Moreover, any subgroup of one of the groups given in Table~\ref{ta:cyclic} also gives rise to a solvable manifold. This means, we can realize any group $\mathbb{Z}_n$ with $n=2,3,4,5,6,8,10,12$.
All of these cyclic subgroups are finite subgroups of the five-dimensional flat group (\ref{5dgroup}).
For all these cases, a Scherk-Schwarz reduction to five dimensions on this flat group is possible. The quantization condition is in five dimensions again given by \eqref{quantphys}, but now with $n=2,3,4,5,6,8,10,12$.

In higher dimensions, the classification of lattices becomes more and more difficult. The reason for the existence of an exhaustive classification in up to five dimensions is the knowledge of finite Abelian subgroups of $SL(n_f)$ for $n_f = 2,3,4$. A similar classification in higher dimensions would give an exhaustive list of quantization conditions for the charges in a Scherk--Schwarz compactification. We will not attempt to do a full classification for dimension six. Let us however remark that, in addition to the cases that come from lower dimensions and are only dressed by some extra vectors $M_a \neq 0$,  in six dimensions we find  new flat  solvmanifolds that correspond to the group $ISO(2)\times ISO(2)$. These  are generalizations of the product of two three-dimensional solvmanifolds with extra shifts $M_a$. Therefore, we do not expect the quantization conditions for each $ISO(2)$ factor to change from \eqref{quantphys} with $n=2,3,4,6$, as long as the solvmanifold is kept  parallelizable.

The classification of flat groups for Scherk-Schwarz reductions generalizes to higher dimensions in a straight-forward way. Moreover, the physical quantization conditions should always be of the form \eqref{quantphys}. In particular, if the charge quantization is similar to \eqref{quant2}, we expect the manifold to be a torus.

\section{Conclusions}
The above considerations illustrate that a consistent truncation to left-invariant modes does not need to coincide with the low-energy effective action of a generic group or coset manifold compactification. The reason for this is that the left-invariant modes might not be the lightest set of fields.
Similarly, the gauged supergravity analysis for the number of preserved supersymmetries of a higher-dimensional solution will only give a lower bound on the actual number of supersymmetries since the Killing spinors do not need to be left-invariant. The proposal of Scherk and Schwarz \cite{Scherk:1979zr} to use flat group manifold compactifications to break supersymmetry and stabilize moduli therefore necessarily requires inhomogeneous solvmanifolds $\mathcal{M} = G/\Gamma$, for which the lattice does not commute with the generators of the group, and for which the left-invariant forms coincide with the basis of de Rham cohomology. As we discussed here, such constructions are based on particular choices of lattice, or equivalently of the quantization of the structure constants of the flat group.
All flat groups  admit also a lattice that does commute with the group generators. The resulting compact space is homogeneous, but is a torus. In these cases, the   supersymmetry breaking and moduli stabilization in the reduced gauged supergravity are an artifact of the truncation, and the ``stabilized moduli'' are simply massive Kaluza-Klein states of the torus.

We have discussed here the flat group compactifications only. Similar considerations apply to more general compactifications with fluxes whenever the internal space is obtained by a lattice action on a group manifold $G$ for which the Lie-group and de Rham cohomologies differ.
On the contrary, for $G/\Gamma$ where $G$ is completely solvable such a problem does not arise and one can  hope that the gauged supergravity captures all massless modes and all unbroken supercharges in a vacuum.

Alternatively, the truncation  to left-invariant modes can be realized as a consistent projection within string theory, by imposing certain boundary conditions in the compact directions  and ``disturbing the symmetry'' between the worldsheet bosons and fermions, as has been studied in \cite{Rohm}. However, this in general leads to theories whose supergravity limit might be different than those obtained by Scherk--Schwarz reduction.

\section*{Acknowledgments}
We thank Carlo Angelantonj, Vicente Cort\'es, Sergio Ferrara, Diego Marques, Diederik Roest, Mario Trigiante, Daniel Waldram, Nick Warner and Fabio Zwirner and specially Massimo Porrati for useful discussions. TVR is supported by a Pegasus Marie Curie fellowship of the FWO.
This work was supported in part by the ERC Starting Grants 259133 -- ObservableString and 40210 - String-QCD-BH and by Agence Nationale de la Recherche under grants 08-JCJC-0001-0 and 12-BS05-003-01.

\bibliographystyle{utphysmodb}
\providecommand{\href}[2]{#2}\begingroup\raggedright\endgroup

\end{document}